**Can the 1.5 deg C warming target be met in a global transition to 100% renewable energy?**

Peter Schwartzman[1] and David Schwartzman[2]
1 Department of Environmental Studies, Knox College, Galesburg, Illinois, USA,
pschwart@knox.edu
2 Department of Biology, Howard University, Washington, DC, USA,
dschwartzman@gmail.com, dschwartzman@howard.edu

**Abstract**

Assuming the state-of-the-science estimates for the remaining carbon budget of carbon dioxide emissions, our modeling study supports the conclusion that it is still possible to meet the 1.5 deg C warming target even with current wind/solar energy technologies, if the creation of a global 100% renewable energy transition of sufficient capacity begins very soon, coupled with aggressive negative carbon emissions.

**Introduction**

In an historic report, the IPCC/WMO (2018) strongly backed the goal of keeping warming at no more than 1.5 deg C, recognizing that meeting this target is still possible, requiring deep and early carbon emissions reductions and obviously far-reaching and unprecedented changes in all aspects of society. Further, this report recognized that implementation of negative carbon emissions technologies is imperative, coupled with radical and early reductions in carbon emissions.  It is increasingly evident that warming above the 1.5 deg C limit will significantly increase the potential for onset of tipping points leading to catastrophic climate change (Lenton et al., 2019).  Since this IPCC report, the latest assessments point to an even greater challenge to keep warming at the 1.5 deg C target (Lenton et al., 2019; Carbon Tracker, 2020). In their review paper, Hilaire et al. (2019) argue that a rapid large-scale deployment of negative carbon emission technologies is imperative to have any chance of achieving the 1.5 deg C target.

Therefore, we examine how this target could still be met, building on our previous simulations (Schwartzman and Schwartzman, 2011).  Our new modeling examines the creation of a global 100% renewable energy supply with a calculation of the cumulative carbon dioxide emitted in a complete phaseout of fossil fuels in a 20 year transition with an assumed starting point at the beginning of 2018. Our objective is to see if estimates of the remaining carbon budget for 1.5 deg C warming are exceeded in our simulations. In addition, we examine whether our computed cumulative carbon dioxide emissions can be reduced at or below this carbon budget using renewable energy-powered negative carbon emission technology, supplemented by potential natural carbon sinks such as restoration of coastal wetlands (Strassburg et al., 2020).

**The Carbon Budget for the 1.5 deg C Warming Target**

Here is a useful definition of a carbon budget for purposes of this discussion:

> A carbon budget is a single number that encapsulates the finite limits of our planet's physical system and highlights the need to reach net zero – if we continue to release emissions on a net basis, the budget is breached and the temperature keeps rising. (Carbon Tracker, 2020).

The world's remaining emissions budget for a 50:50 chance of staying within 1.5 °C of warming is only about 500 gigatonnes (Gt) of $CO_2$. Permafrost emissions could take an estimated 20% (100 Gt $CO_2$) off this budget (ref. 10), and that's without including methane from deep permafrost or undersea hydrates. If forests are close to tipping points, Amazon dieback could release another 90 Gt $CO_2$ and boreal forests a further 110 Gt $CO_2$ (ref. 11). With global total $CO_2$ emissions still at more than 40 Gt per year, the remaining budget could be all but erased already (Lenton et al., 2019, p. 594). (References cited above: 10. Rogelj et al., 2019; 11. Steffen et al., 2018).

On the issue of non-energy contributors to the carbon budget, we note the explanation of Carbon Tracker (2020):

A central difference between IEA and IPCC carbon figures is that whereas the IPCC budgets relate to total emissions of CO2, the IEA scenarios focuses primarily on the energy sector, which is the largest single source of anthropogenic CO2 emissions through the burning of coal, oil and gas. The remainder is accounted for by emissions from industries, land use, land-use change and forestry (LULUCF)… LULUCF alone accounted for 13% of global $CO_2$ emissions in 2018 [iv], so tweaking these assumptions can have considerable implications for the final 'budget' figure. [iv: https://www.globalcarbonproject.org/carbonbudget/index.htm]

Carbon Tracker (2020, Figure 3) shows the IPCC carbon budget for 1.5 deg C as 580 $CO_2$ GT (50% chance), with a start at the beginning of 2018, with a 15% emissions reduction produced in 2018-19 resulting in 493 $CO_2$ GT remaining. In 2020, we witness a significant decline in $CO_2$ emissions as a result of the global COVID-19 crisis (Le Quere et al., 2020). For the start at the beginning of 2018 subtracting 13% for the LULUCF emissions gives an energy-only $CO_2$ budget of 504.6 GT.

The issue of non-$CO_2$ greenhouse contributors should be noted:

Other greenhouse gases (such as methane, fluorinated gases or nitrous oxide) and aerosols and their precursors (including soot or sulphur dioxide) affect global temperatures. Estimating the remaining carbon budget thus also implies making assumptions about these non-$CO_2$ contributions. This further complicates the relationship between future $CO_2$ emissions and global warming (Rogelj et al., 2019, p. 335).

Thus, we conclude that burning natural gas (NG) to produce $CO_2$ is included in the $CO_2$ carbon budget, but the $CO_2$-equivalent warming from methane as a greenhouse gas is not.

**Modeling Results and Discussion**

Modeling computations were done numerically based on the equations shown in Table 1, supplementing the approach of Schwartzman and Schwartzman (2011).

**Table 1.  Equations and Definitions of Model Parameters**

Equations:
1a.  $P_{RE} = (f)^{-1}(F_{ff})(P_{2018})[e^{[(f)(M/L)(t)]} - 1]$
(Modified from equation (2) from Schwartzman and Schwartzman, 2011)

1b.  $R^* = P_{FF} + P_{RE}$

2.  $\Sigma \ CO_2\text{-equivalent emissions} = \Sigma \ E_x \cdot F_x$

Definitions:

$R^*$ is global primary energy consumption relative to the consumption in 2018 corresponding to 18.95 TW in power units (p. 47, IEA, 2020); the renewable component grows to 100% in 20 years as the fossil fuel contributions decline, finally to zero in our simulations.

$P_{FF}$ is the global primary energy consumption of non-renewable energy at time t

$P_{RE}$ is the global primary energy consumption of renewable energy at time t

$P_{2018}$ is the 2018 global primary energy consumption

$F_{RE}$ (f in equation 1) is the fraction of Renewable Energy (wind, solar, CSP) invested to make more of itself

$F_{ff}$ is the fraction of the non-renewable energy supply equal to *present* primary energy consumption level ($P_{2018}$) invested to make Renewable Energy Technology
(constant amount invested over the 20 year transition to 100% renewable supply)

M is the Energy Return/Energy Invested Ratio (EROI) for the Renewable Energy Technology

L is the Lifetime of the Renewable Energy Technology

$E_x$ is the energy consumed annually by each energy source

$F_x$ is the $CO_2$-equivalent emission factor for each energy source

$CO_2$-equivalent [GT] is the $CO_2$ emissions produced in the assumed 20 year energy transition to 100% renewable supply

In these simulations, the renewable energy supply ($P_{RE}$) grows because it is being created from modest contributions of the non-renewable energy component ($F_{ff}$ x $P_{2018}$) along with contributions from the continuously growing renewable supply itself ($F_{RE}$ x $P_{RE}$). The $CO_2$ emissions are computed by summing up the emission factor ($F_x$) multiplied by each energy source as a function of time ($E_x$) (see Equation 2). The assumed EROI ratios equal to 20-30 for the renewable energy technology are consistent with estimates for currently available wind and photovoltaics (Schwartzman and Schwartzman, 2011; Leccisi et al., 2016; Raugei et al., 2012; Raugei and Leccisi, 2016; DeepResource, 2017; Brockway et al., 2018; Vestas, 2019; Rana et al., 2020). As Brockway et al. (2018) emphasize, these renewable technologies now can provide final stage EROI ratios of 20 to 30 for energy (electricity) entering the economy, now even higher than comparable ratios for fossil fuels. We take note of the analysis of comparative EROI ratios of Raugei (2019), with its argument for the plausibility of a renewable energy transition. The advantages of this transition, coupled with boosts in energy conservation and efficiency, also point to the increased availability of supplementary energy needed for energy storage and grid modernization. It is very likely that the EROI ratio of renewable energy will increase in the near future with ongoing research and development, in particular with high efficiency thin-film photovoltaics and floating offshore multi-megawatt wind turbines coming online, making possible a faster renewable energy transition. Modest contributions to the future global energy supply will likely come from geothermal sources used for heating buildings and powering negative carbon emission technology in locales with high heat flow, along with hydropower.

Several simulations test the sensitivity of the assumed constants, with the same computed cumulative $CO_2$ emissions of 423 GT (Figures 1, 2, 3, 4a). Figures 5a and 5b show the assumed phaseouts of the fossil fuels (oil, coal, NG = natural gas) as well as the other energy sources (nuclear, biomass, hydropower and geothermal). Note that oil (conventional) is completely phased out at the end of the assumed transition time of 20 years, while coal and natural gas linearly decrease to zero consumption in 10 years, with one simulation in 20 years (Figure 4b), the same as for the non-fossil fuels energy sources, as the renewable energy supply steadily grows. The slower phaseout of oil was assumed for this fossil fuel with the lowest emission factor in order to supplement the global renewable energy supply as well as remain to the end as a fossil fuel supply creating new renewable energy capacity (the $F_{ff}$ input). The baseline energy consumption of each contributor and assumed emission factors are shown in Table 2. Table 3a and 3b show the computed $CO_2$-equivalent and $CO_2$ only contributions for each fossil fuel for the two "Scenarios" shown in Figures 4a and 4b respectively. Note that the $CO_2$-equivalent and $CO_2$-only emissions are both below the IPCC budget of $CO_2$ energy-only of 505 GT for a January 1, 2018 baseline for the complete phaseout of coal and natural gas in 10 years (Figures 5a and 5b). Nevertheless, the additional potential subtractions from this budget must be taken very seriously (Lenton et al., 2019), hence the imperative to begin this renewable energy transition very soon, coupled with aggressive negative carbon emissions as soon as the renewable energy supply is sufficient for mitigation.

**Table 2 Energy Components and their Emission Factors**

|  | 2018 World Energy Consumption [mtoe] | $CO_2$-equivalent emission factor ($CO_2$ only) [tonnes/TJ] |
|---|---|---|
| Oil (conventional) | 4586 (A) | 76 (68) (B) |
| Coal | 3838 (A) | 106 (90) (B) |
| Natural Gas (NG) | 3262 (A) | 110 (50) (B) |
| Nuclear | 707 (A) | 3 (D) |
|  |  |  |
| Hydropower | 362 (A) | 37 (F) |
| Biomass | 1327 (A) | 60 (E) |
| Geothermal | 7 (C) | 34 (D) |
|  |  |  |
| Wind + Solar | 282 (A) | 3 or 0* (D) |

*3 when generated using fossil fuels, 0 when generated using existing renewable energy.

Sources:
(A) IEA, 2020 ("Other" in source includes wind, solar and geothermal, so geothermal, found from source (C), was subtracted from "Other" ); (B) Howarth, 2020; (C) IRENA, 2017; (D) Pehl et al., 2017; (E) Daioglou et al., 2017; (F) Almeida et al., 2019.

**Table 3a: Total Emissions after 20 years [GT]: Scenario I, No Coal & NG after 10 yrs**

|  | Oil | NG | Coal | Total from FF | Total all Sources |
|---|---|---|---|---|---|
| $CO_2$-equiv. | 200 | 83 | 93 | 375 | 423 |
| $CO_2$-only | 179 | 38 | 79 | 295 | 342 |

**Table 3b: Total Emissions after 20 years [GT]: Scenario II, No Coal & NG after 20 yrs**

|  | Oil | NG | Coal | Total from FF | Total all Sources |
|---|---|---|---|---|---|
| $CO_2$-equiv. | 200 | 158 | 179 | 536 | 583 |
| $CO_2$-only | 179 | 72 | 152 | 402 | 448 |

The amount of oil consumed in our scenarios for the complete transition in 20 years (Figure 5b) is 62,800 mtoe, or 3.6% of the proven global oil reserves of 1,740,000 mtoe (Richie and Roser, 2020). For Scenario II, with full phaseout in 20 years, the total natural gas consumption is equivalent to 35,600 mtoe or 2.0% of proven natural gas reserves and for coal, the total is equivalent to 38,900 mtoe or 5.2% of proven reserves (Richie and Roser, 2020).

Further, while there would be huge health benefits in the elimination of anthropogenic aerosol emissions driven by fossil fuel combustion in a 100% wind/solar transition, cancelling out their significant cooling effect of the order of 0.5 deg C must be taken into account in the estimation of the carbon budget for the 1.5 deg C target (Hausfather, 2018; Zheng et al., 2020). While this effect is included in the IPCC carbon budget, there remains uncertainty in its impact (Hausfather, 2018).

Recycling and industrial ecologies powered by wind/solar power should greatly reduce the need for mining. Recycling rates of the rare earth metals, including neodymium used in wind turbines, is currently very low. Alternative technologies not using rare elements are being developed and implemented, e.g., Na-S batteries, instead of the lower abundant lithium. Hence, a transition to a post-extractivist future is possible, promoted by a wind/solar transition, which should be coupled with a strong regulatory regime necessary for environmental, worker and community protection.

Further, the scenario outlined here would very likely require progressive demilitarization of global society (Lorincz, 2014), i.e., the phasing out of the Military- Industrial Complex, which would liberate vast quantities of materials, especially metals, for the creation of a global wind and solar power infrastructure. Likewise, it would free up materials in the conversion of the automobile-roadways complex to electrified rail and public transit powered by wind-solar energy sources. And last but not least, demilitarization will promote a badly needed international regime of cooperation so critical to a rapid transition elimination the global consumption of fossil fuels (Schwartzman and Schwartzman, 2019). We take note of the potential synergy between demilitarization, termination of fossil fuel consumption and the improvement of the quality of life, especially in the global South, as a result of the freeing up of vast material and financial resources; direct and indirect costs of the global subsidization of fossil fuels amount to $5 trillion a year (Coady et al., 2019) while the global annual military expenditures are on the order of $2 trillion (SIPRI, 2020).

**The role of negative carbon emissions to meet the 1.5 deg C target**

The most energy-efficient approaches to mitigation of carbon dioxide emissions are switching to renewable energy combined with energy efficiency measures in industry, transportation and buildings (Babacan et al., 2020). This mitigation needs to be coupled with negative carbon emissions, which is getting a lot of attention by the scientific community following the IPCC/WMO (2018) report, with increasing focus on Direct Air Capture and Storage (DACCS) technologies (Beuttler et al., 2019; Realmonte et al., 2019; Chatterjee and Huang, 2020; Rosa et al., 2020). A promising approach is Climeworks-CarbFix injection into subsurface basalt at Hellisheidi, Iceland (Beuttler et al., 2019; Snæbjörnsdóttir et al., 2020).

In their review of mineral carbonation as a process for carbon sequestration from the atmosphere, Snæbjörnsdóttir et al. (2020) cites a range of the energy requirement of 3.4 to 10.7 GJ/ton of $CO_2$ captured from air, with the upper limit corresponding to 471 kJ/mole $CO_2$. Assuming a rough requirement of 500 kJ/mole $CO_2$ for Direct Air Capture (DAC) leading to permanent storage via subsurface mineral carbonation, to reduce the carbon dioxide emissions by 150 GT in 10 years would require the use of the equivalent of 5.38 TW renewable energy production. Reducing the energy requirement to the lower limit of 3.4 GJ/ton of $CO_2$ cited by Snæbjörnsdóttir et al. (2020)

would correspond to 1.61 TW of renewable energy production or for a drawdown of 300 GT in 10 years, 3.22 TW. These energy estimates are well within the computed total renewable production created by 2028 in simulations with $F_{RE} \geq 0.15$. In addition, some of the necessary energy could apparently come from other sources linked to the mineral carbonation process:

> The efficiency of the DAC process is inevitably directly linked with the $CO_2$ emissions generated by its energy source; for maximum efficiency, the energy demand of the DAC process should be supplied by available low-grade industrial-waste heat or low-$CO_2$-emitting geothermal heat (Snæbjörnsdóttir et al., 2020; p. 98).

These rates of carbon storage, from 15 to 30 GT $CO_2$/year, are consistent with the higher estimates for potential sequestration in the crust for air capture in onland peridotite (Kelemen et al., 2019, figure 4).

Of course, in these simulations we are not arguing for the need to create a renewable energy infrastructure with a primary energy consumption of 2 to over 3 times the present consumption level in 20 years. Rather, taking into account technological improvements in $2^{nd}$ law (exergy) efficiencies reducing the necessary global energy demand, the goal should likely be below 1.5 times the present level, to have the capacity to eliminate global energy poverty, for climate mitigation and adaptation as well as other challenges posed in the next few decades, thereby optimizing the quality of life of all of humanity along with the preservation of biodiversity. Jacobson et al. (2017) project a 23% improvement in efficiency from 100% renewably-driven electrification by 2050.

Several well-researched studies have concluded that the necessary energy demand in a renewable energy transition could be significantly lower than the present without the need for direct air capture technologies, rather relying on restoration of natural ecosystems (Jacobson et al., 2017, 2019; Grubler et al., 2018; Millward--Hopkins et al., 2020). Recognizing the importance of this research, we conclude that their level of projected energy demand is insufficient to eliminate energy poverty in several high population countries of the global South such as India, as well as to effectively meet the challenges of climate mitigation and adaptation (Schwartzman and Schwartzman, 2019).

A rough calculation can provide some idea of future energy needs in 20 to 30 years from now. Assuming a minimum of 3 kW/person minimum for achieving the highest world standard life expectancy, 9 billion people and an efficiency gain of 30% for a 100% renewable energy transition, 3 x 9 x 0.70 = 19 TW (in power units) which is equal to the present global primary energy consumption. Adding incremental needs, climate mitigation and adaptation, biosphere cleanup and ecosystem restoration will of course increase this estimate. There is significant uncertainty in just how much more energy in the future will be needed globally to confront climate change, highly contingent on future warming scenarios (van Ruijven et al., 2019). Just considering the energy demands of climate mitigation in our scenarios, at least the increment of 3 to 11 TW, would be needed. Not including the drawdown from restoration of natural ecosystems which could significantly reduce this requirement, we take this increment as a minimum since other incremental needs are not included.

We recognize the debate between the pessimists (Smil, 2010, 2016) and optimists (Diesendorf and Elliston, 2018; Jacobson et al., 2019) regarding the possibility of making such a rapid energy transition. Jacobson et al. (2019) argue:

> The roadmaps call for countries to move all energy to 100% clean, renewable wind-water- solar (WWS) energy, efficiency, and storage no later than 2050 with at least 80% by 2030 (Abstract).

Smil (2016) concludes:

> Replacing the current global energy system relying overwhelmingly on fossil fuels by biofuels and by electricity generated intermittently from renewable soemiurces will be necessarily a prolonged, multidecadal process (Abstract).

We note that Jacobson et al. (2019)'s roadmap does not invoke the need for biofuels while making the case for a reliable baseload energy supply by renewable sources. Diesendorf and Elliston (2018) conclude:

> We find that the principal barriers to 100RElec are neither technological nor economic, but instead are primarily political, institutional and cultural" (Abstract).

We agree, with the biggest barrier being the existing political economies on our planet.

This scenario of rapid transition to wind/solar energy should be coupled with aggressive energy conservation in energy-wasteful countries in the global North such as the U.S. thereby making possible a rapid buildup of renewable energy production in the global South to terminate energy poverty. Randers and Goluke (2020) argue that self-sustained temperature rise from runaway melting of permafrost could potentially be avoided by atmospheric removal of 33 GT $CO_2$-equivalent per year. Recognizing that the modeling and conclusions of this study have been vigorously challenged by climate scientists (Hanley, 2020), the renewable energy requirement for this flux is consistent with the potential generation in our modeling for $F_{RE} \geq 0.15$.

Of course, negative carbon emission technologies can be significantly supplemented by the restoration of natural ecosystems, such as grasslands/prairies, coastal wetlands and forests, thereby providing a carbon sink from the atmosphere into biomass and soil. For example, an estimate of 299 GT of $CO_2$ could be sequestered with this approach (Strassburg et al., 2020). However, industrial scale carbon sequestration from the atmosphere into the crust will be necessary for both short term and long term time scales since the carbon sink of natural ecosystems will saturate and diminish their capacity even with an additional 0.5 deg C of warming, as the oceans will continue to reequilibrate with the atmosphere, resulting in a carbon dioxide flux to the atmosphere as this gas is sequestered into the crust. Indeed, Realmonte et al. (2019) point out:

> From our scenarios, DACCS is foreseen to remove between 16 and 30 $GtCO_2$/year over the period 2070–2100. This implies that a significant fraction (from 10 to 19%) of the carbon removed would be released back to the atmosphere from the oceans, requiring an additional removal of 1.7 to 9.5 GtCO$_2$/year to meet the same carbon budget (p. 8).

Water and material requirements for infrastructure must be addressed in a robust program of negative carbon emissions (Chatterjee and Huang, 2020; Rosa et al., 2020). The use of sea water in subsurface mineral carbonation (Kelemen et al., 2019; Carbfix, 2020) is likely the best option, along with solar-powered synthesis of chemicals used in DAC, such as monoethanolamine, from air-derived nitrogen and carbon (Luis, 2016; Sefidi and Luis, 2019), although more efficient technologies can be anticipated in the future.

Carbon capture and sequestration (CCS) from fossil fuel power plants (Eldardiry and Habib, 2018) increasingly powered by the renewable energy infrastructure should arguably be a component of a plan to optimize the rapid reduction of carbon emissions while providing the necessary global energy supply, but taking note of the important critique of Jacobson (2019) we emphasize that this mode of CCS should not be implemented to prolong the unnecessary consumption of fossil fuels.

**Conclusion**

It is technically possible to keep warming at or below the 1.5 deg C warming goal even utilizing the present state-of-the-science renewable energy technologies in a 20 year transition to 100% global renewable energy supply with a rapid termination of fossil fuels with the highest emission factor. As many have already recognized, the challenge is overcoming the political economic obstacles to meet this goal, in order to begin this transition in the near future.

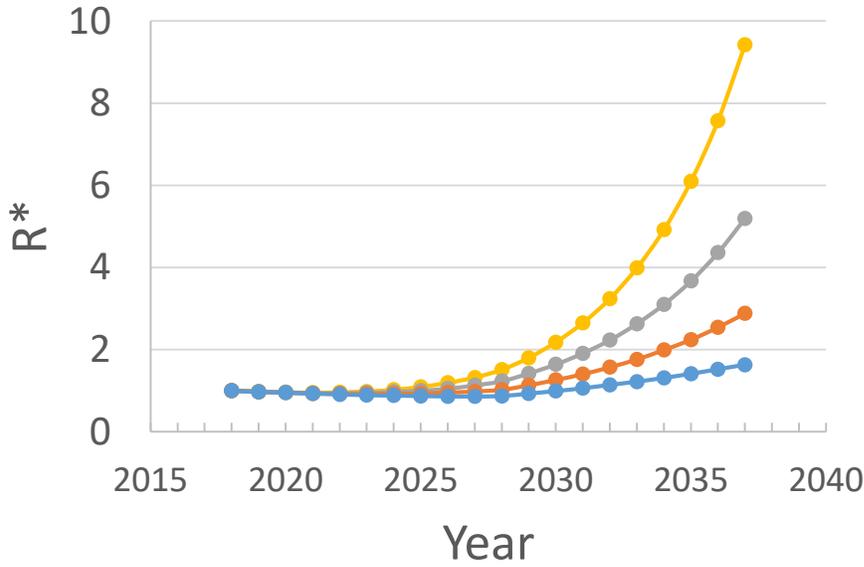

Figure 1: Future Energy Consumption with varying RE reinvestments ($F_{RE}$)

$F_{ff} = 0.03$
$M = 20$, $L = 20$
CO2-equiv. [GT] = 423

$F_{RE} = 0.25$
$F_{RE} = 0.20$
$F_{RE} = 0.15$
$F_{RE} = 0.10$

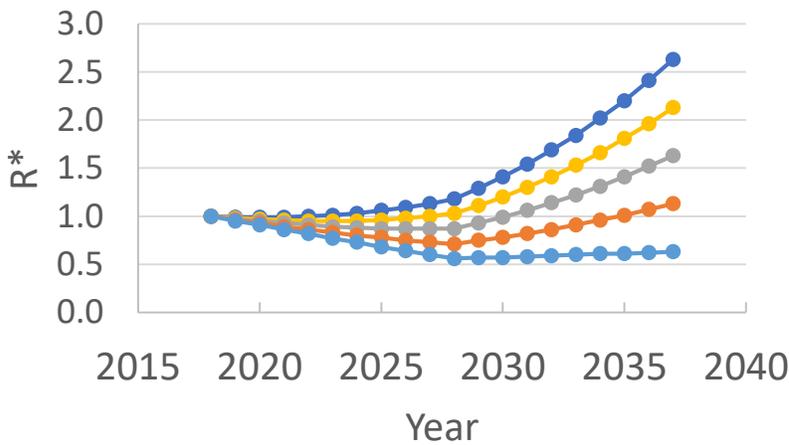

Figure 2: Future Energy Consumption with varying fossil fuel reinvestments ($F_{ff}$)

$F_{RE} = 0.10$
$M = 20$, $L = 20$
CO2-equiv. [GT] = 423

$F_{ff} = 0.05$
$F_{ff} = 0.04$
$F_{ff} = 0.03$
$F_{ff} = 0.02$
$F_{ff} = 0.01$

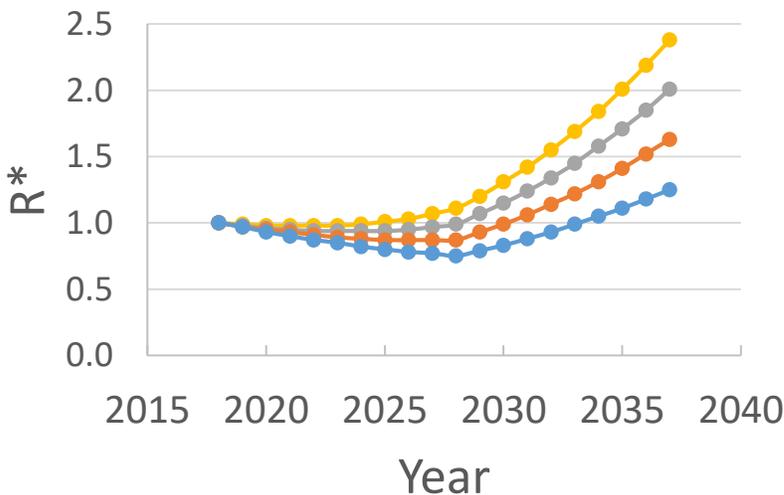

Figure 3: Future Energy Consumption with varying M

$F_{ff} = 0.03$, $F_{RE} = 0.10$
$L = 20$
CO2-equiv. [GT] = 423

M = 30
M = 25
M = 20
M = 15

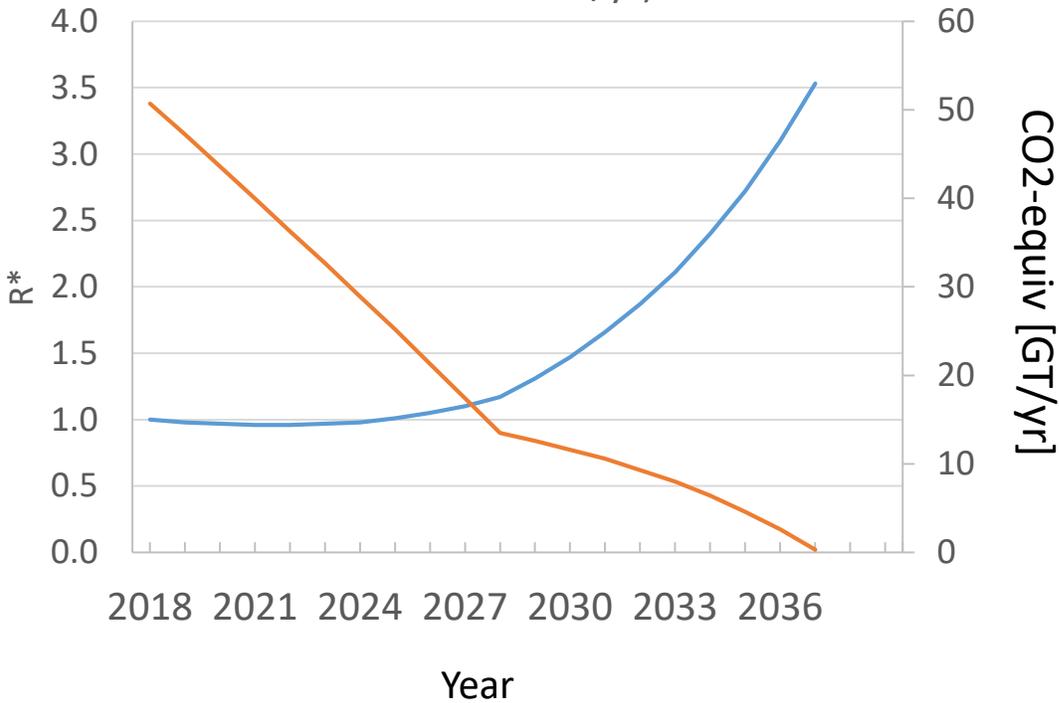

Figure 4a:
Scenario I – Coal & NG end in 10 yrs
(Energy Consumption, CO2-equiv. emissions/yr)

$F_{ff}$ = 0.03, $F_{RE}$ = 0.15
L = 20, M = 25
CO2-equiv. [GT] = 423

— R*
— CO2-equiv. [GT]/yr

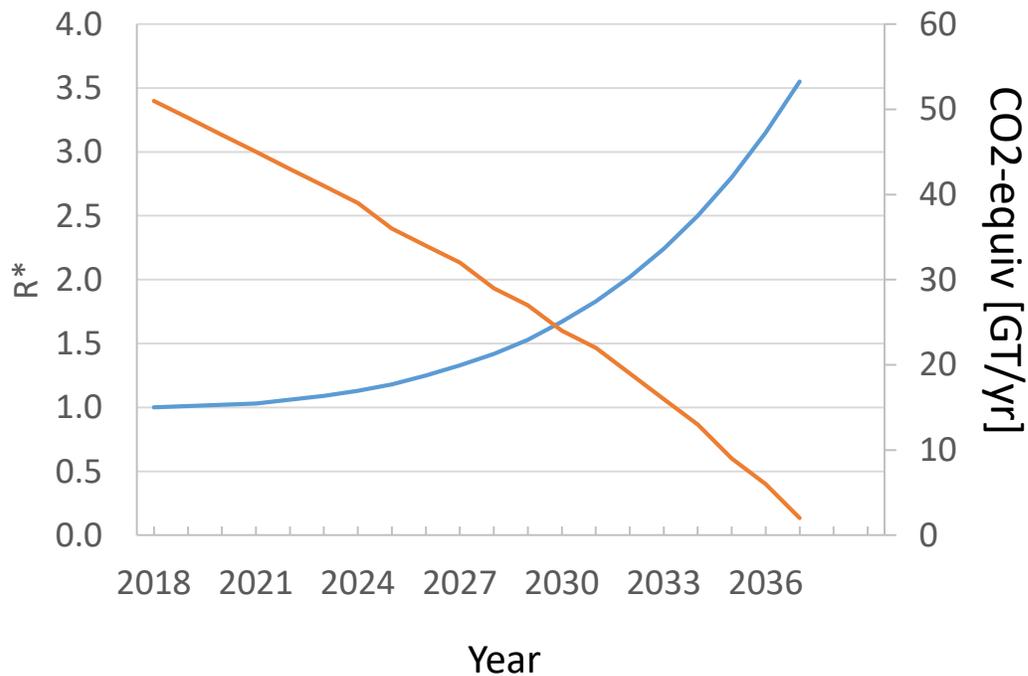

Figure 4b:
Scenario II—Coal & NG end in 20 yrs
(Energy Consumption, CO2-equiv. emissions/yr)

$F_{ff}$ = 0.03, $F_{RE}$ = 0.15
L = 20, M = 25
CO2-equiv. [GT] = 583

— R*
— CO2-equiv. [GT]/yr

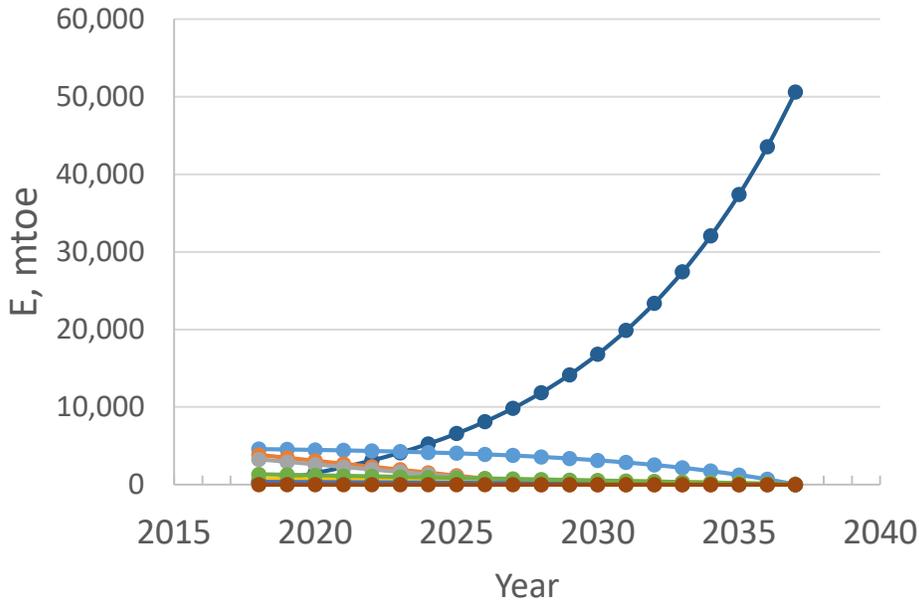

Figure 5a: Energy Consumption by Form w/RE
(Scenario I, No Coal & NG after 10 yrs)

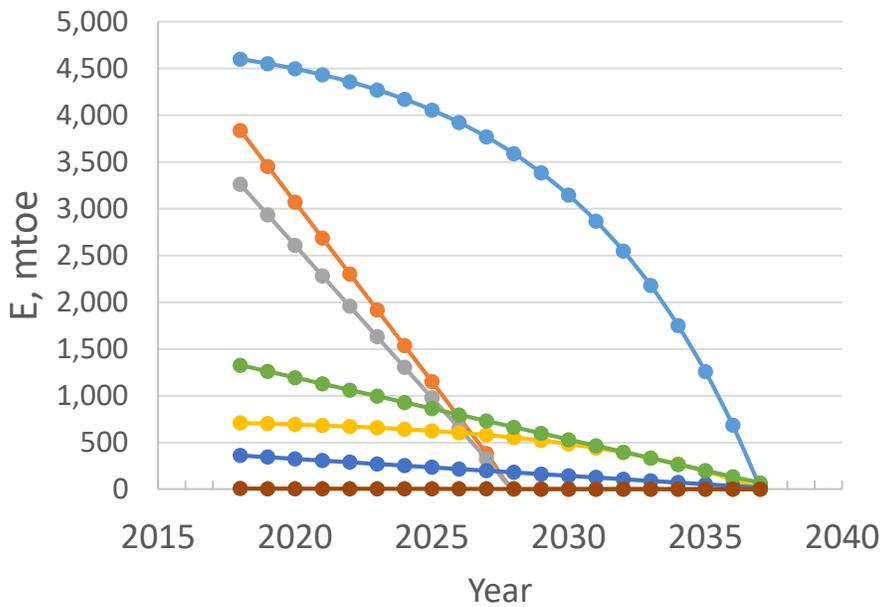

Figure 5b: Energy Consumption by Form
(Scenario I, No Coal & NG after 10 yrs)

LEGEND for Figures 5a & 5b:

At 20 yrs:
CO2-equiv. [GT] = 423
CO2 [GT] = 342

- RE = Wind + Solar
- Oil
- Coal
- NG
- Nuclear
- Hydro
- Biomass
- Geothermal

Note: 10,000 mtoe = 13.27 TWy